\documentclass[twocolumn]{aastex61}

\usepackage{color}
\usepackage{epsfig}
\usepackage{amsmath}
\usepackage{enumerate}
\usepackage{epstopdf}
\usepackage{verbatim}
\usepackage{courier}
\usepackage[flushleft]{threeparttable}
\usepackage{tabularx}
\usepackage{gensymb}

\begin{document}

\title{On the detection of coronal dimmings and the extraction of their characteristic properties}
\correspondingauthor{K. Dissauer}
\email{karin.dissauer@uni-graz.at}

\author{K. Dissauer}
\affiliation{Institute of Physics/IGAM, University of Graz, 8010 Graz, Austria}
\author{A. M. Veronig}
\affiliation{Institute of Physics/IGAM, University of Graz, 8010 Graz, Austria}
\affiliation{Kanzelh\"ohe Observatory/Institute of Physics, University of Graz, 9521 Treffen, Austria}
\author{M. Temmer}
\affiliation{Institute of Physics/IGAM, University of Graz, 8010 Graz, Austria}
\author{T. Podladchikova}
\affiliation{Skolkovo Institute of Science and Technology, 143026 Moscow, Russia}
\author{K. Vanninathan}
\affiliation{Institute of Physics/IGAM, University of Graz, 8010 Graz, Austria}

\begin{abstract}
Coronal dimmings are distinct phenomena associated to coronal mass ejections (CMEs). The study of coronal dimmings and the extraction of their characteristic parameters helps us to obtain additional information of CMEs, especially on the initiation and early evolution of Earth-directed CMEs. We present a new approach to detect coronal dimming regions based on a thresholding technique applied on logarithmic base-ratio images.
Characteristic dimming parameters describing the dynamics, morphology, magnetic properties and the brightness of coronal dimming regions are extracted by cumulatively summing newly dimmed pixels over time. 
It is also demonstrated how core dimming regions are identified as a subset of the overall identified dimming region. 
We successfully apply our method to two well-observed coronal dimming events. For both events the core dimming regions are identified and the spatial evolution of the dimming area reveals the expansion of the dimming region around these footpoints. We also show that in the early impulsive phase of the dimming expansion the total unsigned magnetic flux involved in the dimming regions is balanced and that up to 30\% of this flux results from the localized core dimming regions. Furthermore, the onset in the profile of the area growth rate is co-temporal with the start of the associated flares and in one case also with the fast rise of the CME, indicating a strong relationship of coronal dimmings with both flare and CMEs. 
\end{abstract}


\section{Introduction} \label{sec:intro}
Coronal dimmings are regions of reduced extreme-ultraviolet (EUV) and soft X-ray (SXR) emission in the low corona \citep{Hudson:1996,Sterling:1997,Thompson:1998, Thompson:2000} that occur in association with CMEs. Their appearance is in general interpreted as density depletion caused by the evacuation of plasma during the CME lift-off \citep{Hudson:1996,Thompson:2000,Harrison:2000}.
Simultaneous and co-spatial observations of coronal dimmings in different wavelengths \citep{Zarro:1999} as well as plasma outflows identified from spectroscopic observations support this interpretation \citep{Harra:2001,Attrill:2010a,Tian:2012}. Also studies on dimming/CME mass relations \citep{Sterling:1997,Wang:2002,Harrison:2003,Zhukov:2004,Mandrini:2007,Aschwanden:2009,Aschwanden:2016,Lopez:2017} and Differential Emission Measure (DEM) studies indicating localized density drops in dimming pixels by up to 70\% (Vanninathan et al. 2017) provide further evidence for this interpretation.

Two different types of dimmings can be identified, core (or twin) dimmings and secondary (or remote) dimmings, respectively. Core dimming regions are stationary, localized regions, close to the eruption site, and rooted in opposite magnetic polarity regions. In simple configurations they are interpreted to mark the footpoints of the ejected flux rope \citep{Mandrini:2005,Mandrini:2007, Muhr:2010}. Secondary dimmings are more shallow and diffuse and can extend to significant distances from the eruption site developing in all directions symmetrically or in narrow, extended structures \citep{Chertok:2005}. They are sometimes observed as rarefaction regions to follow behind a propagating EUV wave front \citep[e.g.][]{Muhr:2011}.
Here, we focus on secondary dimmings related to the CME expansion, that are most likely formed by the plasma evacuation behind the flux rope and overlying fields that are erupting.

The detection of coronal dimming regions is not trivial, since dimmings are different in geometry and level of intensity decrease, show a non-uniform development and parts of their area can be covered by flare brightening. In literature different approaches are used to identify the location of coronal dimming regions.  
\cite{Reinard:2008} use base-difference maps to detect dimming pixels. All pixels that are 1$\sigma$ below the mean value of the whole differenced image are detected as part of the dimming. They characterize each dimming region by its area (number of dimming pixels) and brightness (sum over their intensity). 
A similar approach is discussed in \cite{Attrill:2010b}.
\cite{Podladchikova:2005} developed the Novel EIT wave Machine Observing (NEMO) algorithm, which is based on the analysis of the general statistical properties of eruptive on-disk events. The dimming detection is also performed on base-difference images, in which two sets of pixels are identified, applying a higher and lower sensitive threshold. The final dimming is extracted using a region growing algorithm, starting from the 1\% darkest pixels and growing into the region identified with the lower sensitive threshold, fulfilling the condition of a simply-connected region.
\cite{Kraaikamp:2015} use also a combination of low and high thresholding to extract coronal dimming regions. Their input consists of running difference (RD) and percentage running difference images (PRD). They obtain their final dimming mask by only keeping clusters from the RD mask that intersect with clusters in the PRD mask. \cite{Krista:2013} perform the detection of coronal dimming pixels on direct images using a local intensity thresholding method on Lambert cylindrical equal-area projection SDO/AIA \mbox{193 \AA}~maps. 

Solar EUV images are usually logarithmically scaled since the intensities from the different regions can vary over orders of magnitude. When quantifying the changes in such images over time, this concept allowing more clearly to present the results, should still be considered.
Therefore, in this paper, we present a new approach for the detection of coronal dimming regions and also the extraction of their characteristic parameters. It is based on a thresholding algorithm that is applied to logarithmic base-ratio images. In the first part of the paper a detailed description of the method is given. Section~\ref{sec:detection} illustrates why base-ratio images favor the detection of coronal dimming regions, especially in quiet Sun regions and how the core dimming regions are extracted as a subset of the initially identified overall dimming region. 
In Section~\ref{sec:parameters} the concept of cumulative dimming masks, representing the full extent of the total dimming region over time is introduced. Based on these masks, characteristic coronal dimming parameters, describing the dynamics, morphology, magnetic properties and the brightness of the dimmings are derived. 
In the second part of the paper we apply our new approach to two well-observed coronal dimming events.
The first event occurred on September 6, 2011 and is used throughout the paper to illustrate the detection and the extraction of parameters. In Section~\ref{sec:results} the results for this event and the second event that occurred on December 26, 2011 are presented. It is well studied in literature \citep{Cheng:2016,Qiu:2017} and acts therefore as a test case for the application of our new approach. In addition to the properties of the total dimming regions (i.e.~covering both core and secondary dimmings), also the spatial extent and the magnetic fluxes of the core dimming regions are given. We discuss advantages of our method, as well as the main findings of the two cases under study in Section~\ref{sec:discussion}.
\section{Data and Pre-processing} 
We use high-cadence (12~s) data from seven different extreme-ultraviolet  (EUV) wavelengths of the Atmospheric Imaging Assembly (AIA; \citealt{Lemen:2012}) on-board the Solar Dynamics Observatory (SDO; \citealt{Pesnell:2012}) covering a temperature range of \mbox{$\approx 50,000- 1.0\times 10^{7}$~K}. The 171, 193, \mbox{211 \AA}~channels represent plasma at quiet Sun temperatures (0.6--2 MK), while 335, 94 and \mbox{131 \AA}~are sensitive to hot plasma of active regions and flares (up to 10 MK). The temperature response function of the \mbox{304 \AA~}filter peaks at \mbox{$\approx$ 50,000} K~and plasma at this temperature is likely to origin from the transition region and chromosphere.

To study the magnetic properties of coronal dimming regions, the 720~s line-of-sight (LOS) magnetograms of the SDO/Helioseismic and Magnetic Imager (HMI; \citealt{Scherrer:2012,Schou:2012}) are used. The data is rebinned to \mbox{$2048 \times 2048$} pixels under the condition of flux conservation. Standard Solarsoft IDL software is used to prepare the data (\mbox{\texttt{aia\_prep.pro}} and \mbox{\texttt{hmi\_prep.pro}}). We check for constant exposure time and exclude AIA images where the automatic exposure control algorithm was triggered. We correct for differential rotation by rotating each data set to a common reference time using \mbox{\texttt{drot\_map.pro}}.
Furthermore, we restrict the detection of coronal dimming regions to a subfield of \mbox{$1000\times 1000$~arcsecs} around the center of the eruption.

To study the time evolution of coronal dimming events our time series covers 12 hours, starting 30~minutes before the associated flare. For the first two hours the full-cadence (12~s) observations of SDO/AIA are used, while for the remaining time series the cadence of the observations is successively reduced to 1, 5, and 10~minutes, respectively. This setting allows us to study the impulsive phase of the dimming in great detail as well as its recovery phase later on.
\section{Coronal dimming detection}\label{sec:detection}
To study the global properties of coronal dimmings, we first identify the total dimming region, including both core and secondary dimming regions. In addition, we identify core dimming regions as a subset of the overall dimming, in order to obtain more information on the distribution of the different types within the total dimming region. 
\subsection{Logarithmic base-ratio images}
\begin{figure*}
\centering
\includegraphics[width=0.8\textwidth]{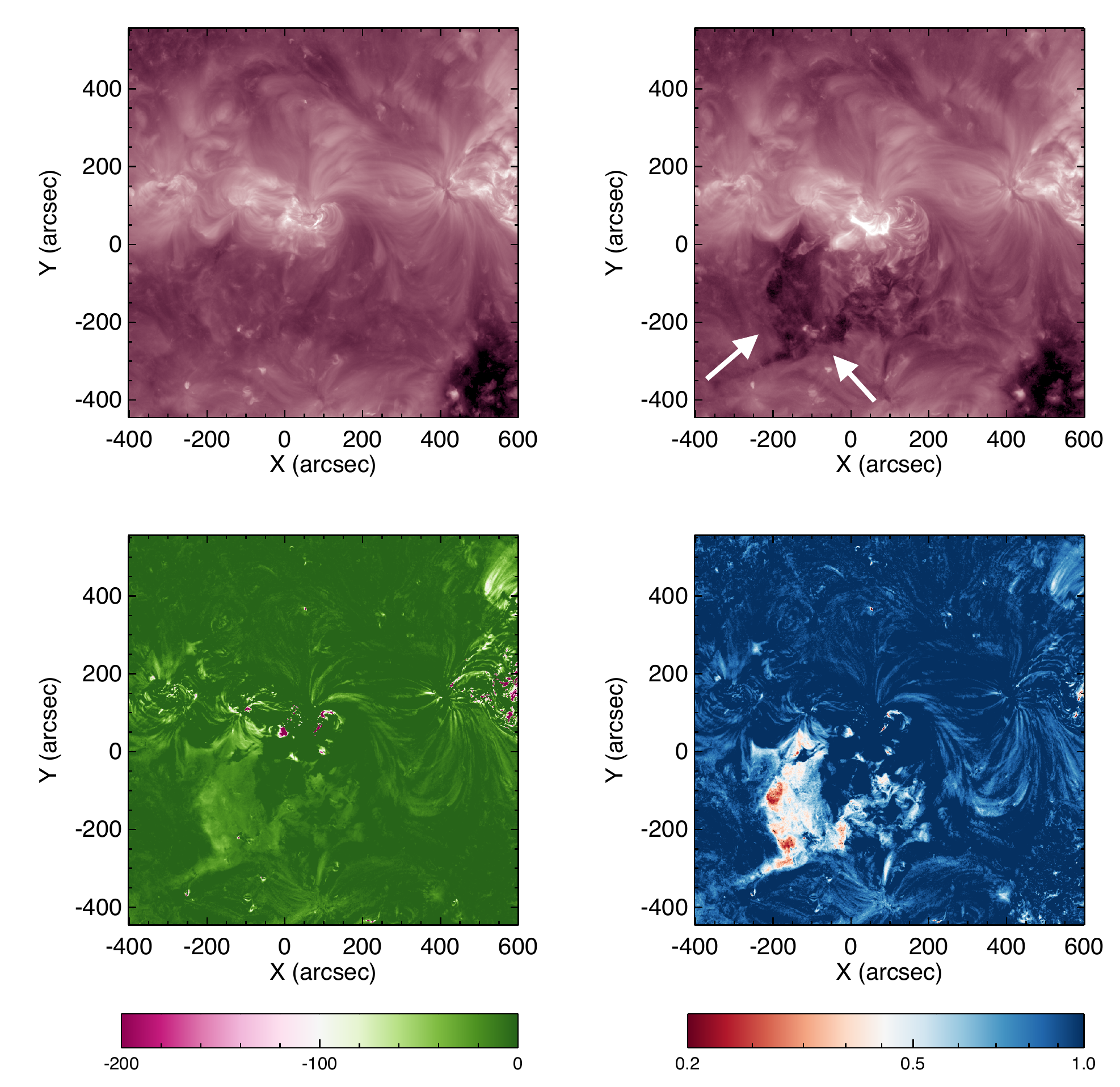}
\caption{Example illustrating that logarithmic base-ratio images are in favor for the detection of coronal dimming regions compared to base-difference images. Top panels: Logarithmically scaled SDO/AIA \mbox{211~\AA}~filtergrams before the start of the eruption (left) and at the maximal extent of the coronal dimming region on October 1, 2011 (right). Bottom panels: The corresponding base-difference (left) and logarithmic base-ratio images (right) are calculated from the top panels. The maximum scaling of the images is set to the level of no change (0.0 for base-difference and 1.0 for base-ratio images). Regions that increase in intensity over time are no longer visible since they are set to the maximum scaling. Dimming regions, i.e. regions that decrease in intensity are more pronounced and better visible. The white arrows indicate a coronal dimming region observed from the direct observations.}
\label{fig:comparison}
\end{figure*}
One important issue in extracting coronal dimming regions is the type of data representation used for their detection. 
In general, base-difference images are used to identify the changes in intensity of coronal dimming regions. To construct these images, a pre-event reference image is subtracted from each subsequent frame. Base-difference images favor the visibility of absolute intensity changes, this means regions of large intensity changes will be restricted to active regions and coronal loop regions, while the simultaneous detection of dimmed regions in the quiet Sun may be problematic. 
Figure~\ref{fig:comparison} shows an example illustrating this issue. The top panels show SDO/AIA \mbox{211~\AA}~filtergrams close to the maximum extent of a coronal dimming region (right) together with a pre-event reference image taken 30~minutes before the start of the eruption (left). The direct observations show the expansion of a dimming region towards the South-East (indicated by white arrows).
The lower left panel shows the corresponding base-difference image calculated from the top panels. Regions of strong intensity decrease occur close to the eruption center and in nearby active regions. The dimming region shows only moderate changes in intensity, at a similar level of small-scale fluctuations. 
In this example, base-difference images used for the detection will not be able to properly identify the coronal dimming region although it is even seen in the direct images.

Therefore, we propose to use the relative change in intensity (base-ratio images) to identify dimming regions. Each image is divided by a pre-event frame, so changes in low intensity regions are equally weighted compared to changes in high intensity regions. 
For even better visibility we use logarithmically scaled base-ratio images, since they map the intensity of strongly decreased (``dark") pixels over a wider intensity scale compared to linear scaling.
Such representation of solar EUV images is generally used, as the intensities from the different regions vary over several orders of magnitude. 
We calculate logarithmic base ratio images 
\begin{equation}
LBR=\log\left(\frac{I_{n}}{I_{0}}\right)=\log(I_{n})-\log(I_{0}) \;,
\end{equation}
with $I_{n}$ and $I_{0}$ being the intensities of the image at time $t_{n}$ and at the reference time $t_{0}$ for the base image. Note that the logarithmic base ratio images are equivalent to base-differences of the logarithmically scaled images.
The application of logarithmic base-ratio images is illustrated in the lower right panel of Fig.~\ref{fig:comparison}, where the dimming region can be easily identified. Regions with moderate intensity decrease appear from light blue to white, while regions of strong intensity decrease ($\geq$50\%) are shown from white to red. Regions that did not show an intensity change or increased in intensity (e.g. due to the flare) are shown in dark blue.
\subsection{Detecting coronal dimming pixels}\label{sec:detect_pixel}
\begin{figure*}
\centering
\includegraphics[width=0.95\textwidth]{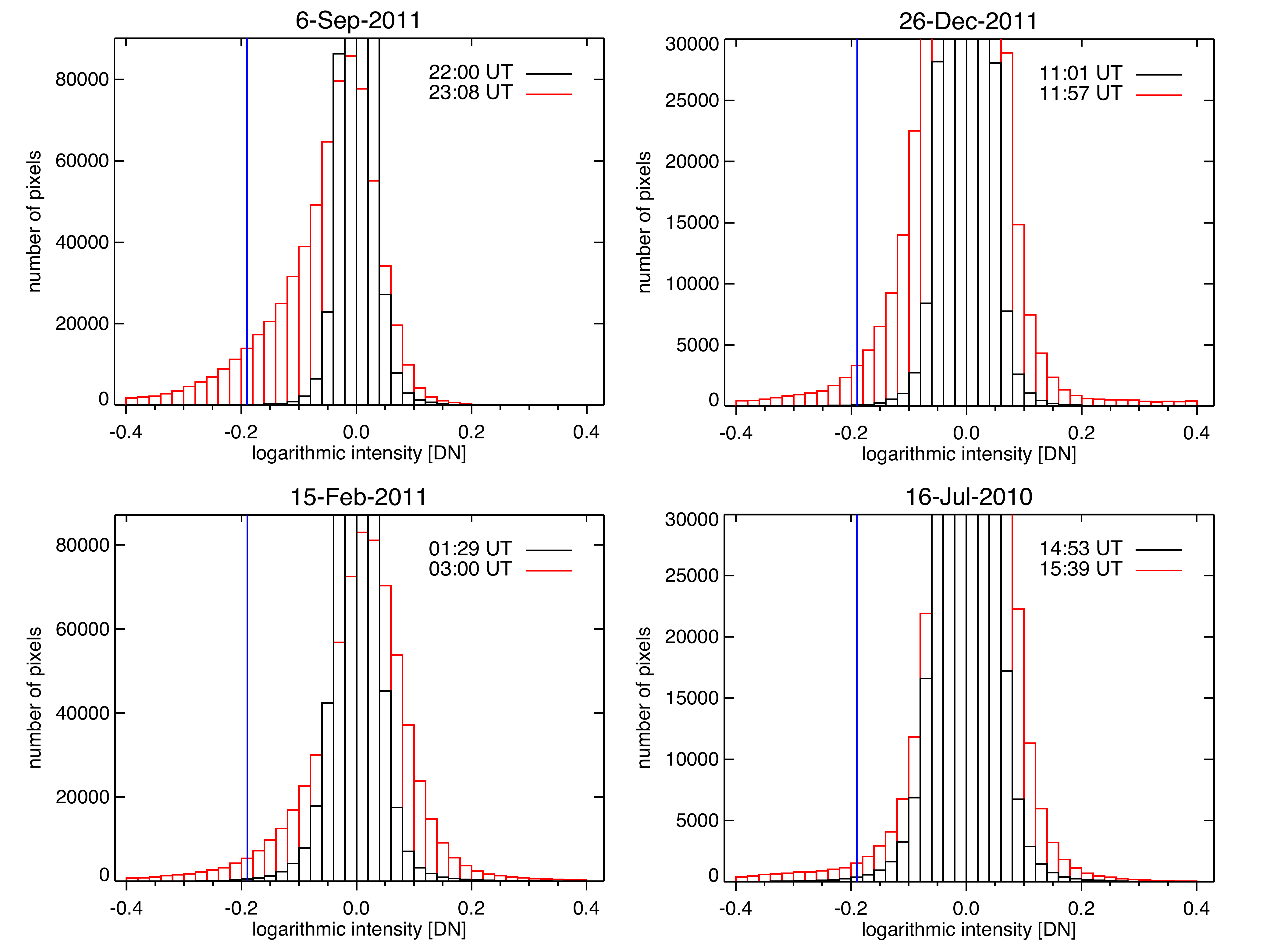}
\caption{Intensity distributions of logarithmic base-ratio images at the non-dimming state (black) and at the dimming state (red) for four events, representing strong and weak dimming events. While the black distribution shows small-scale changes prior to the dimming expansion, the red histograms show a significant shift towards negative values and therefore the existence of dimming pixel populations. The vertical blue line marks the threshold used for the dimming detection, which lies below the noise level and allows robust detection of pixels that belong to the coronal dimming. (Note that for clearer visibility of the differences in the distributions, we confined the y-range).}
\label{fig:threshold}
\end{figure*}

To identify coronal dimming pixels, we apply a thresholding algorithm on logarithmic base-ratio images.
A pixel is flagged as a dimming pixel when its logarithmic ($\log_{10}$) ratio intensity dropped below \mbox{-0.19} DN, which corresponds to a percentage change of about 35\% in linear space.
This threshold was determined empirically by comparing the intensity distributions calculated from images observed before the dimming formation (non-dimming state) and during its maximal extent (dimming state) for different dimming events. Figure~\ref{fig:threshold} illustrates such histograms for four events, representing strong and weak dimmings. For all events, the non-dimming distributions (indicated in black) cover an intensity range from \mbox{-0.15 to 0.15~DN}. This indicates that small-scale changes and fluctuations exist already prior to the formation of dimmings, e.g. due to the displacement of coronal loops. The histograms representing the dimming state (indicated in red) are shifted towards negative values (i.e. towards intensities that decrease) compared to the non-dimming state. These pixel populations represent pixels that are part of the dimming regions.
The vertical blue line marks the threshold used for the dimming detection. It lies below the non-dimming noise level of \mbox{-0.15~DN} but still allows the detection of the real dimming pixels.
As a next step, morphological operators (IDL routines \texttt{dilute.pro} and \texttt{erode.pro}) are applied. These operators are used to smooth the extracted regions, i.e.~small features (e.g. noise) are removed, while small holes and gaps are filled. In this way seed pixels for a region growing algorithm \mbox{(IDL routine \texttt{region\_grow.pro})} are obtained. Finally, only detected pixels that are connected neighbors to the seed pixels are extracted as final dimming regions. In this way, the detection of small-scale fluctuations is reduced and misidentified pixels are removed, especially for lower signal-to-noise filters of AIA (e.g. 94 and \mbox{131 \AA}) and filters where the dimming is only barely visible (\mbox{304 \AA}). 

\begin{figure*}
\centering
\includegraphics[width=0.85\textwidth]{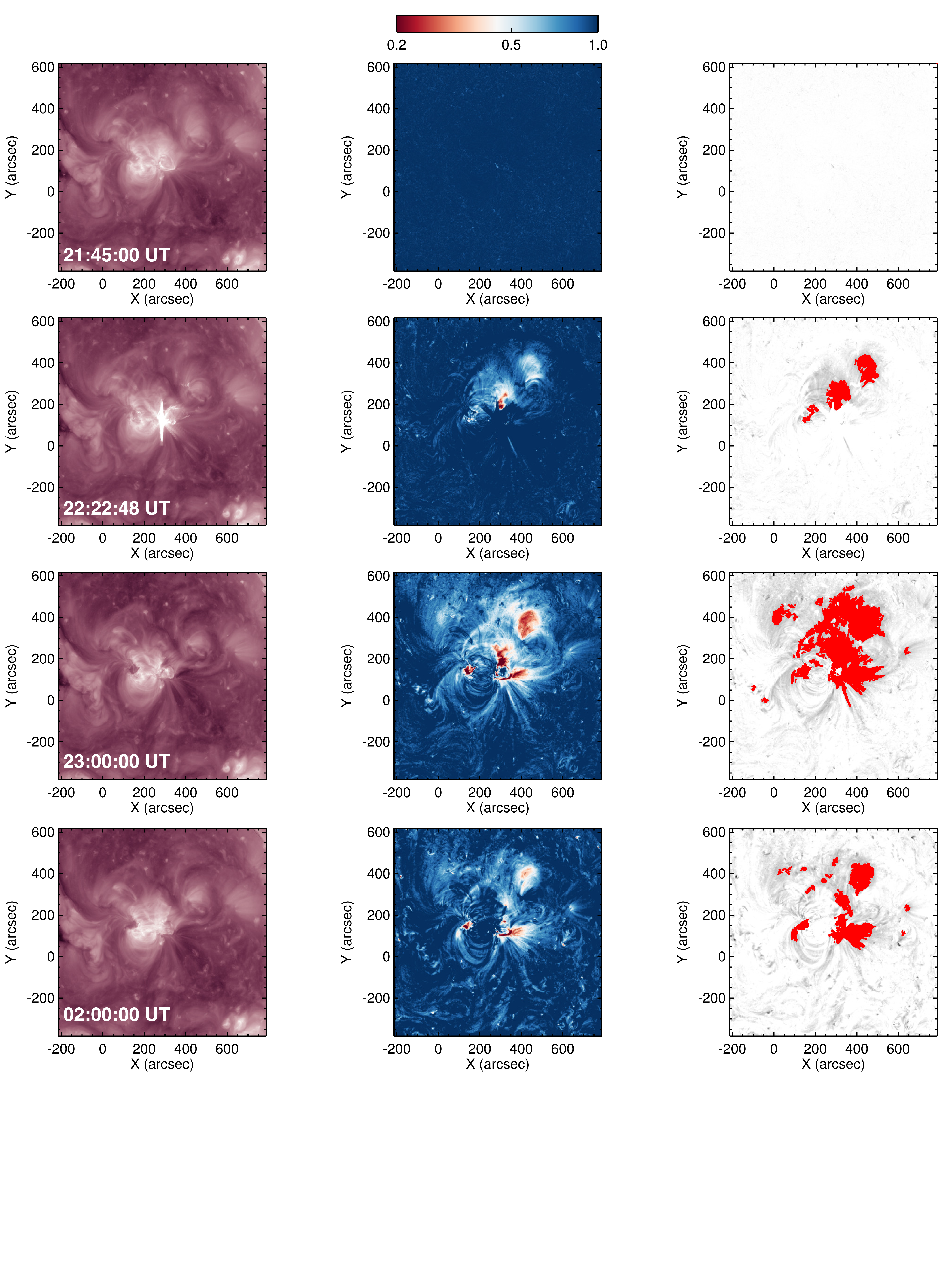}
\caption{Sequence of SDO/AIA 211 \AA~direct (left) and logarithmic base-ratio (middle) images showing the time evolution of the expansion of the coronal dimming on September 6, 2011. The right panels show the identified dimming pixels for each time step in red (see also accompanied animation no.~1).}
\label{fig:detection}
\end{figure*}

Figure~\ref{fig:detection} illustrates the application of the detection procedure on a sequence of SDO/AIA \mbox{211 \AA}~images for the coronal dimming event on 6 September, 2011 associated to a \mbox{X2.1} flare/CME event (see also accompanying animation no.~1). While it is difficult to identify the secondary dimming regions from the direct images (left panels), the logarithmically scaled base-ratio images show the formation of a pronounced dimming region towards the North (middle panels). The identified dimming pixels for each time step are indicated in red in the right panels.

\begin{figure*}
\begin{center}
\includegraphics[width=0.62\textwidth]{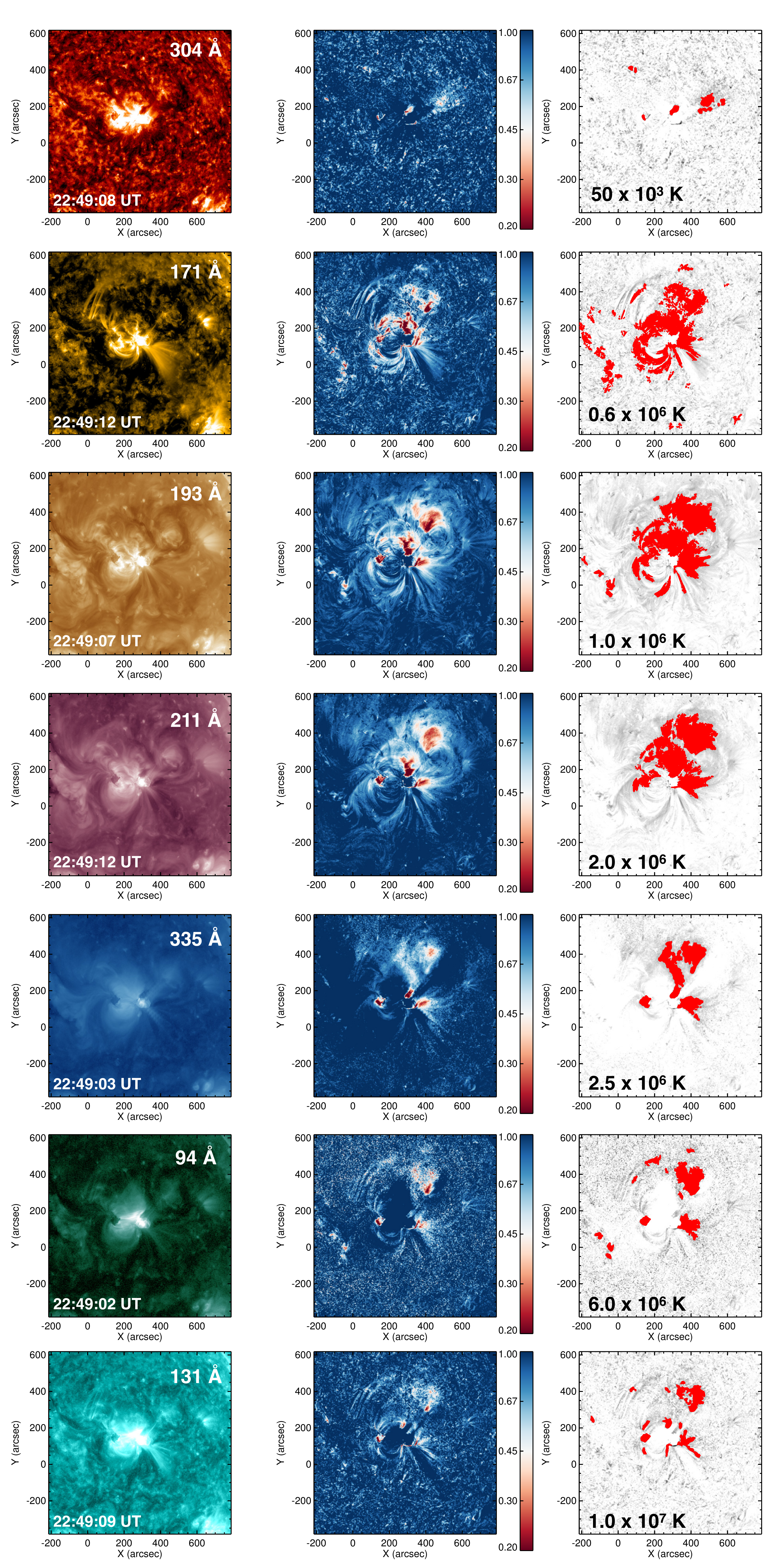}
\caption{Detection of the coronal dimming regions around \mbox{22:49~UT} on 6 September 2011 in seven EUV filters of SDO/AIA covering a temperature range from \mbox{0.5--10~MK} (top to bottom). The direct image (left) is plotted together with its corresponding logarithmic base-ratio image (middle) and the detected dimming pixels (indicated in red in the right panel).}
\label{fig:multi-wavelength}
\end{center}
\end{figure*}

Figure~\ref{fig:multi-wavelength} shows the multi-wavelength application of the detection algorithm for September 6, 2011 at the time of maximum extent of the dimming region. We are able to successfully detect coronal dimming areas in all seven EUV filters of SDO/AIA. The shape and area of the identified dimming regions changes for different wavelengths indicating that plasma at different temperatures is evacuated. For this event, the largest dimming regions are observed in the 211, 193 and \mbox{171 \AA}~filters. This means that the coronal dimming regions are mostly formed by the evacuation of plasma at quiet Sun coronal temperatures around \mbox{1--2} MK.

\subsection{Core dimming}\label{sec:core}
Core dimming regions should be observed close to the eruption site in regions of opposite magnetic polarity. 
In addition spectroscopic and DEM studies showed signatures of strong outflows of dense plasma as well as a strong decrease in density and intensity \citep[][Vanninathan et al. 2017]{Tian:2012}. 
We assume potential core dimming regions to be a subset of the total dimming region and we use minimum intensity maps (cf.~Sect.~\ref{sec:brightness}) of base-difference and logarithmic base-ratio data to identify these regions. Each core dimming pixel fulfills the following two conditions: (1) it is detected in the early impulsive phase of the total dimming (within the first 30 minutes after the onset)  and (2) its pixel intensity decreased below both thresholds A and B, calculated from minimum intensity maps of base-difference and logarithmic base-ratio data, respectively. These thresholds are defined as follows
\begin{equation}
\begin{split}
A&=\bar{I}_{BD}-0.6\;\sigma_{BD} \;, \\
B&=\bar{I}_{LBR}-0.6\;\sigma_{LBR} \;,
\end{split}
\end{equation}
where $\bar{I}_{BD}$ is the mean intensity of the base-difference minimum intensity map, $\bar{I}_{LBR}$ is the mean intensity of the logarithmic base-ratio minimum intensity map, $\sigma_{BD}$ and $\sigma_{LBR}$ are the corresponding standard deviations.
This means that we identify the darkest dimming regions in terms of both absolute and relative intensity change. 

\begin{figure*}
\includegraphics[width=1.0\textwidth]{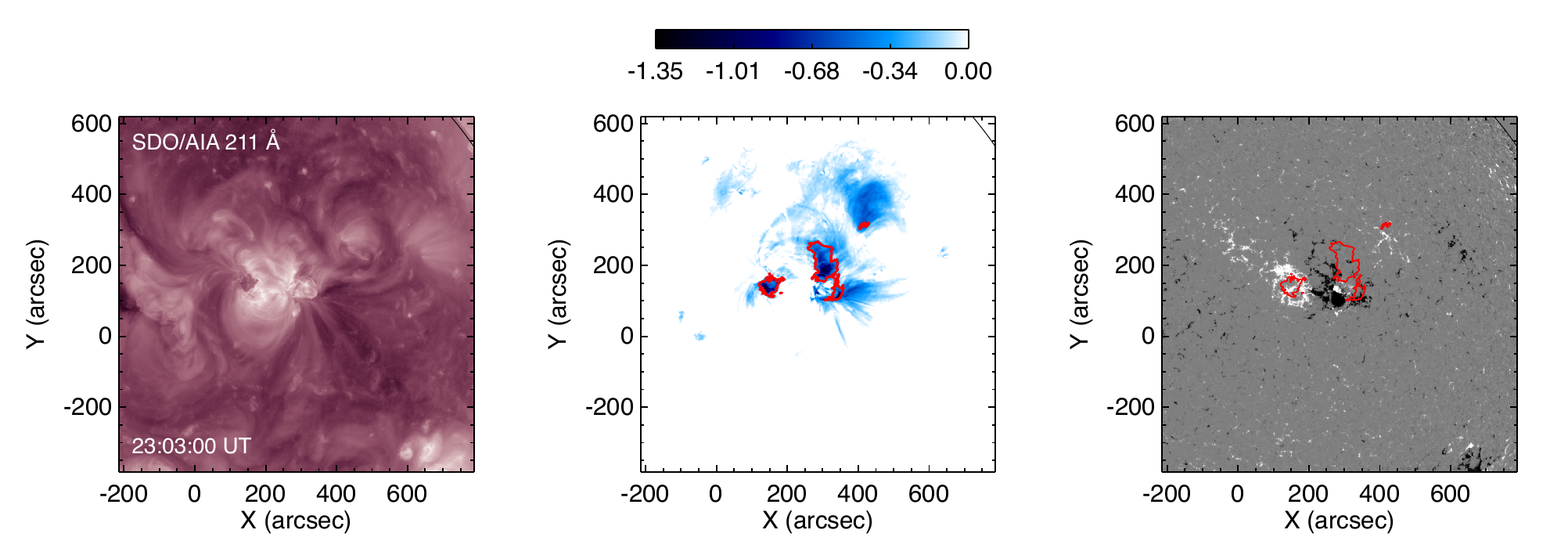}
\caption{SDO/AIA \mbox{211 \AA}~filtergram close to the time of maximum extent of the coronal dimming region (left) together with the minimum intensity map of logarithmic base-ratio data (middle) and the LOS SDO/HMI magnetogram (right) for the September 6, 2011 event. The identified ``potential" core dimming regions are marked by the red contours. Those regions are located in opposite magnetic polarity regions, close to the eruption site (center of the FOV).}
\label{fig:core}
\end{figure*}

We choose the minimum intensity maps for the detection since plasma is evacuated over a certain period of time and not all pixels might show their minimum intensity simultaneously. Using these maps we make sure to cover all pixels at minimum intensity over time. 
Figure~\ref{fig:core} illustrates the identification of core dimming regions for September 6, 2011. It shows a SDO/AIA \mbox{211 \AA}~filtergram close to the time of the maximum extent of the dimming (left), the minimum intensity map of logarithmic base-ratio data (middle), together with the corresponding LOS magnetogram from SDO/HMI (right). The red contours outline the dimming pixels identified as core dimming regions. Two localized patches located in opposite polarity regions within the active region are identified. They show evacuation of dense plasma, observed as maximum decrease in intensity in the total dimming region. We therefore conclude that these regions  found by our algorithm indeed point to the footpoints of the erupting flux rope.
\section{Characteristic coronal dimming parameters}\label{sec:parameters}
We extract and analyze different characteristic parameters of coronal dimming regions and their time evolution in order to describe their physical properties. These parameters represent the dynamics, morphology, magnetic properties and brightness of the dimming. The calculated parameters reflect the properties of the total dimming region, including both types of dimming, i.e.~core and secondary dimmings.

On the one hand, we analyze quantities that describe the instantaneous state of the dimming. They are determined from dimming regions identified for each time step $t_{i}$, individually, i.e.~instantaneous dimming pixel masks.
On the other hand, we also calculate parameters that describe the dynamics of the phenomenon better by using all dimming pixels that are identified over a certain \textit{time range}, i.e.~cumulative dimming pixel masks.
We define these pixel masks as follows.
For each time step $t_{m}$ we identify coronal dimming pixels $p_{i}$ and define an \textit{instantaneous} dimming pixel mask $D_{m}(p_{i}, t_{m})$ with a value of 1 at the locations of identified dimming pixels and 0 elsewhere. This mask represents all pixels identified as dimming pixels for a certain time step $t_{m}$. A \textit{cumulative} dimming pixel mask $M(p_{i},t_{n})$ is then created over time by the union of instantaneous dimming pixel masks identified from $t_{0}$ until $t_{n}$, i.e.
\begin{equation}
\begin{split}
M(p_{i}, t_{n})&=D_{n}(p_{i}, t_{n})\;\cup\;D_{n-1}(p_{i}, t_{n-1})\;\cup \ldots \\
&\cup\;D_{0}(p_{i}, t_{0})\;.
\end{split}
\end{equation}
$M(p_{i},t_{n})$ contains then every pixel that was flagged as a dimming pixel between time $t_{0}$ and $t_{n}$ \citep[see also][]{Kazachenko:2017}.

\begin{figure*}
\centering
\includegraphics[width=1.0\textwidth]{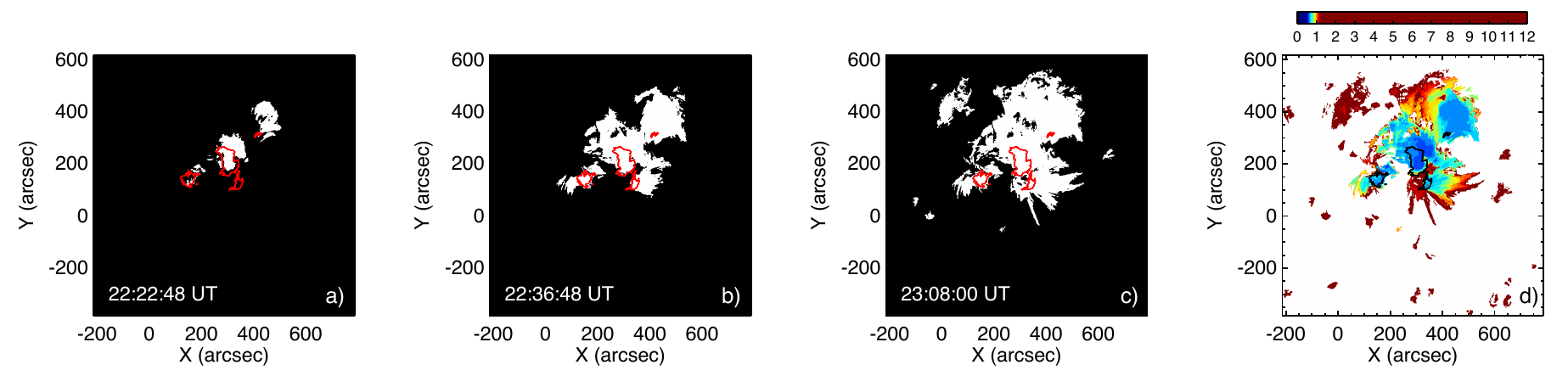}
\caption{Evolution of cumulative dimming pixel masks representing all dimming pixels that are detected during the main impulsive phase of the coronal dimming (until \mbox{23:08~UT}). Right: Corresponding timing map of the coronal dimming region indicating the time at which a pixel is detected as a dimming pixel for the first time (color coded in hours after the flare onset). The majority of dimming pixels are detected within the first two hours of the time series. The red and black contours mark the identified core dimming regions.}
\label{fig:mask_timing}
\end{figure*}

Figure~\ref{fig:mask_timing} shows as an example the evolution of the cumulative dimming pixel mask of the coronal dimming that occurred on 6 September, 2011. The right panel presents the final cumulative dimming pixel mask for the full time range of 12 hours, with each pixel colored by the time of its first detection (in hours after the flare onset). This \textit{timing map} illustrates the spatial evolution of the total dimming region over time.

In the calculation of dimming area and magnetic fluxes, we account for projection effects and correct the area and the LOS magnetic field density for each detected dimming pixel with respect to its position on the solar disk \citep{Hagenaar:2001}. In the following we describe each calculated dimming parameter in more detail, an overview of all extracted quantities is given in  Table~\ref{tab:parameters}.
\renewcommand{\arraystretch}{1.25}
\begin{table*}
\centering
\caption{Overview of the characteristic coronal dimming parameters}
\begin{threeparttable}
\begin{tabularx}{0.7\textwidth}{l l l}
\hline
  $A$ & $=\sum_{i} M(p_{i},t)a_{i}$ &area  \\
  $\dot{A}$ & $=\frac{dA}{dt}$ & area growth rate \\ \hline
  $\Phi_{+}$ &$=\sum_{i}B_{+}(p_{i})M(p_{i},t)a_{i}$  & positive magnetic flux\\
  $\Phi_{-}$ & $=\sum_{i}B_{-}(p_{i})M(p_{i},t)a_{i}$ & negative magnetic flux\\
  $\Phi$ & $=\left(\Phi_{+}+|\Phi_{-}|\right)/2$ & total unsigned magnetic flux\\
  $\dot{\Phi}_{+}$ & $=\frac{d\Phi_{+}}{dt}$ & positive magn. flux rate\\
  $\dot{\Phi}_{-}$ & $=\frac{d\Phi_{-}}{dt}$ & negative magn. flux rate\\
  $\dot{\Phi}$ & $=\frac{d\Phi}{dt}$ & total unsigned magn. flux rate\\
  $\bar{B}$ & $=(\Phi_{+}+\Phi_{-})/2A$ & mean magnetic field density\\ 
  $\bar{B}_{+}$ & $=\frac{\Phi_{+}}{A_{+}}$ & mean pos. magn. field density\\
  $\bar{B}_{-}$ & $=\frac{\Phi_{-}}{A_{-}}$ & mean neg. magn. field density\\ \hline
  $I_{inst}$ & $=\sum_{i} I(p_{i},t) D_{n}(p_{i},t)$ & instantaneous total brightness\\
  $I_{cu}$ & $=\sum_{i} I(p_{i},t) M(p_{i},t)$ &cumulative total brightness\\
  $I_{min}$ & $=\sum_{i} min(I(p_{i},t) M(p_{i},t))$ &minimum total brightness\\\hline
\end{tabularx}
\label{tab:parameters}
\begin{tablenotes}
\item \textbf{Note.} The subscript $i$ denotes a pixel within the dimming pixel masks, $a_{i}$ its area, $B_{i}$ its magnetic field density, $I(p_{i}, t)$ its intensity at time step $t$.
\end{tablenotes}
\end{threeparttable}
\end{table*}
\subsection{Morphological properties}
The morphology of coronal dimmings is studied by extracting the area of all dimming pixels identified until time $t_{n}$
\begin{equation}
A(t_{n})=\sum_{i} M(p_{i},t_{n})a_{i}
\end{equation}
where $a_{i}$ is the area of each dimming pixel $p_{i}$ within the cumulative dimming pixel mask $M(p_{i},t_{n})$. In this way we capture the full combined extent of all dimming pixels until $t_{n}$, i.e. also including dimming regions that have been covered by the flare in the beginning or by post-flare loops later on.
Its time derivative, is given by the area growth rate calculated as
\begin{equation}
\frac{dA}{dt}=\frac{A(t_{n})-A(t_{n-1})}{t_{n}-t_{n-1}}\;.
\end{equation}
The time evolution of both quantities together with the timing map (cf.~right panel, Fig.~\ref{fig:mask_timing}) allows us to study in detail the spatial and temporal morphology of the dimming region.
\subsection{Magnetic properties}
In order to understand the complete magnetic footprint of coronal dimmings we extract separately the positive and negative magnetic flux as well as the total unsigned magnetic flux within the dimming region until $t_{n}$ as follows:
\begin{equation}
\Phi_{+}(t_{n})=\sum_{i}B_{+}(p_{i})M(p_{i},t_{n})a_{i} \;,
\end{equation}
\begin{equation}
\Phi_{-}(t_{n})=\sum_{i}B_{-}(p_{i})M(p_{i},t_{n})a_{i} \;,
\end{equation} 
\begin{equation}
\Phi(t_{n})=\frac{\left(\Phi_{+}+|\Phi_{-}|\right)}{2} \;,
\end{equation}
only using pixels where the LOS magnetic field density exceeds the noise level of SDO/HMI ($|B|>10$~G). The errors in the magnetic fluxes are estimated by by varying the intensity threshold used for the detection of the dimming pixels (see Sect.~\ref{sec:detect_pixel}) by $\pm$ 5\%.
For each parameter we also calculate the time derivative in order to quantify the corresponding magnetic flux change rates.

Finally, we determine the mean magnetic field density $\bar{B}$, the mean positive magnetic field density $\bar{B}_{+}$ and the mean negative magnetic field density $\bar{B}_{-}$ covered by the dimming region until time $t_{n}$ as
\begin{equation}
\bar{B}(t_{n})=\frac{\Phi_{s}}{A}=\frac{(\Phi_{+}+\Phi_{-})}{2 A} \;,
\end{equation}
\begin{equation}
\bar{B}_{+}(t_{n})=\frac{\Phi_{+}}{A_{+}} \;\;\;,\;\;\; \bar{B}_{-}(t_{n})=\frac{\Phi_{-}}{A_{-}} \;,
\end{equation}
where $A_{+}$ and $A_{-}$ is the dimming area with underlying positive and negative magnetic field density, respectively.
The study of these magnetic properties helps us to understand how much magnetic flux is involved in the dimming region and therefore ejected to interplanetary space by the CME, it is balanced and whether it is coming from isolated, magnetically strong regions. 

\begin{figure*}
\centering
\includegraphics[width=0.85\textwidth]{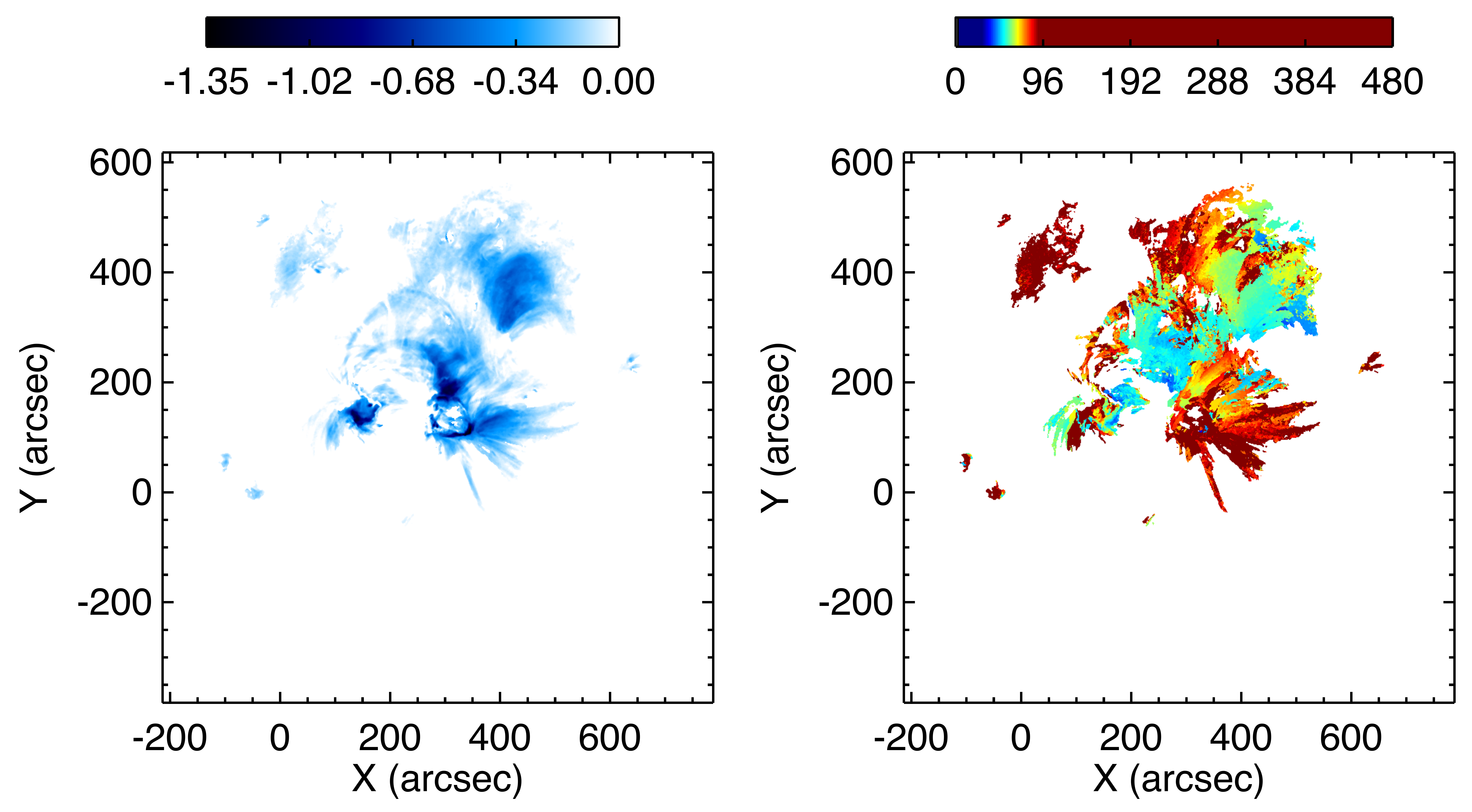}
\caption{Left: Minimum intensity map of logarithmic base-ratio data (left) for SDO/AIA 211\AA~filtergrams on September 6, 2011 (color-coded according to the color bar on top of the panel). Right: Timing map illustrating when each pixel reaches its minimum value (color coded from black to red in minutes after the corresponding flare onset). The majority of pixels is dimmed to its minimum during the first two hours of the time series and the darkest intensity pixels are located close to the eruption site.}
\label{fig:minimum_brightness_timing}
\end{figure*}
\subsection{Dimming brightness}\label{sec:brightness}
If coronal dimming regions represent low coronal footprints of CMEs, the total brightness of dimming regions can be interpreted as a measure of the amount of evacuated plasma and therefore the mass of the CME. The darker and the bigger the dimming region the more plasma should be evacuated \citep{Mason:2016,Aschwanden:2016}. 
In general the total brightness is defined as the sum of dimming pixel intensities using base-difference or base-ratio images. Depending on how these dimming pixels are defined different information from the time evolution of this quantity can be extracted.
To study the instantaneous total brightness $I_{inst}$ of the dimming region at time step $t_{m}$, we calculate
\begin{equation}
I_{inst}(t_{m})=\sum_{i} I(p_{i},t_{m}) D_{m}(p_{i},t_{m})\;,
\label{eq:tot_bright}
\end{equation}
where $I(p_{i},t_{m})$ is the intensity at $t_{m}$ for a certain dimming pixel $p_{i}$ within the instantaneous dimming mask $D_{m}(p_{i},t_{m})$. This quantity allows us to investigate the instantaneous state of the dimming region in terms of intensity and area.
A minimum in the time evolution of $I_{inst}$ can result either from a maximum dimming area or from a minimum intensity of these pixels.
A measure for the total brightness of the dimming region only depending on the intensity change of the dimming pixel is given by
\begin{equation}
I_{cu}(t_{m})=\sum_{i} I(p_{i},t_{m}) M(p_{i},t_{n}) \;,
\end{equation}
where $I(p_{i},t_{m})$ is the intensity at $t_{m}$ for a certain dimming pixel $p_{i}$ within a fixed cumulative dimming mask $M(p_{i},t_{n})$.
In contrast to Eq.~\ref{eq:tot_bright} the dimming area used is constant over the full time range. This area is defined by all dimming pixels that have been identified from start $t_{0}$ to time $t_{n}$.
This definition of the total dimming brightness is in particular suited to study the recovery phase and the replenishment of the corona in the evacuated dimming regions.

Coronal dimmings can extend far from the eruption site, with some parts reaching their lowest emission value long before other regions of the dimming. In order to calculate the total minimum brightness of the dimming over time, it is therefore important to identify the full dimming area at its minimum intensity. To this aim we introduce \textit{minimum intensity maps}. We identify all dimming pixels during e.g. the impulsive phase of the dimming, i.e. all pixels that were detected between $t_0$ and $t_{n}$. For each individual pixel of this final cumulative dimming pixel mask its minimum intensity over time is extracted. This map allows us to calculate the total minimum brightness of the dimming region over the full time range $t$
\begin{equation}
I_{min}=\sum_{i} min(I(p_{i},t)M(p_{i},t_{n}))\;.
\end{equation}
The minimum brightness map also indicates the locations of the darkest pixels over time within the overall dimming region.
A similar concept, known as persistence maps was introduced by \cite{Thompson:2016} and applied to coronal dimming regions. They claim that especially the total mass represented by a dimming, or the footprint of all of the magnetic fields involved in dimmings, would benefit from using such maps. Therefore, we use minimum intensity maps for the identification of core dimming regions representing the footpoints of the erupting flux rope (cf.~Sec.~\ref{sec:core}).

Figure~\ref{fig:minimum_brightness_timing} shows the minimum intensity map of logarithmic base-ratio data for September 6, 2011 (left panel). The darkest pixels are located close to the eruption site. The right panel represents a timing map that illustrates when each pixel of the mask reaches its minimum intensity. It can be seen that the majority of pixels is dimmed to its minimum during the first two hours of the time series.

We restrict the time range to search for the minimum intensity of each pixel to the first five hours of the time series. Changes that occur later are often due to overall variations of the corona and projection effects that occur due to derotation of each image to the base. The time evolution of $I_{cu}$ (Fig.~\ref{fig:profiles}, (e)) further supports this restriction. For the dimming region on September 6, 2011 the relaxation phase of the corona starts already within the first two hours of the time series. \\
\section{Case study results}\label{sec:results}
We characterize in detail the dimming events that occurred on 6 September 2011 and 26 December 2011 based on our new approach. These events are part of a larger statistical sample, all events are selected on-disk from the perspective of SDO within 40\degree~around the central meridian and were associated with Earth-directed CMEs.

The dimming on 6 September 2011 was used throughout the paper to illustrate our new detection method and how dimming parameters are extracted. In the following, the time evolution of the dimming parameters derived are described and interpreted.

The coronal dimming that occurred on 26 December 2011 is one of the rare examples of well studied coronal dimming events using recent data in literature \citep{Cheng:2016,Qiu:2017}. We will discuss the results obtained from our new algorithm in context to findings reported in these papers. 
\subsection{6 September 2011}
The event is associated with a X2.1 flare in NOAA active region 11283, at heliographic position \mbox{N14\degree W18\degree}. The impulsive phase of the flare starts at \mbox{22:16~UT} and reaches its peak around \mbox{22:21~UT}. A globally propagating EUV wave, that is most pronounced towards the North is also observed (average speed \mbox{v=1100~km~s$^{-1}$}, derived from SDO/AIA). The associated CME is detected as a halo CME from the SOHO/LASCO coronagraph with an average speed of \mbox{v=550~km~s$^{-1}$}. The speed as derived from STEREO-A, observing the CME almost in the plane of sky, is much higher (average \mbox{v=990~km~s$^{-1}$}, maximal \mbox{v=1200~km~s$^{-1}$}), indicating the influence of strong projection effects on the kinematics derived from SOHO/LASCO. Also for the analysis of the EUV wave and the dimming region combining the data from different vantage points was essential to correctly interpret these phenomena \citep{Dissauer:2016}.

\begin{figure}
\includegraphics[width=1.0\columnwidth]{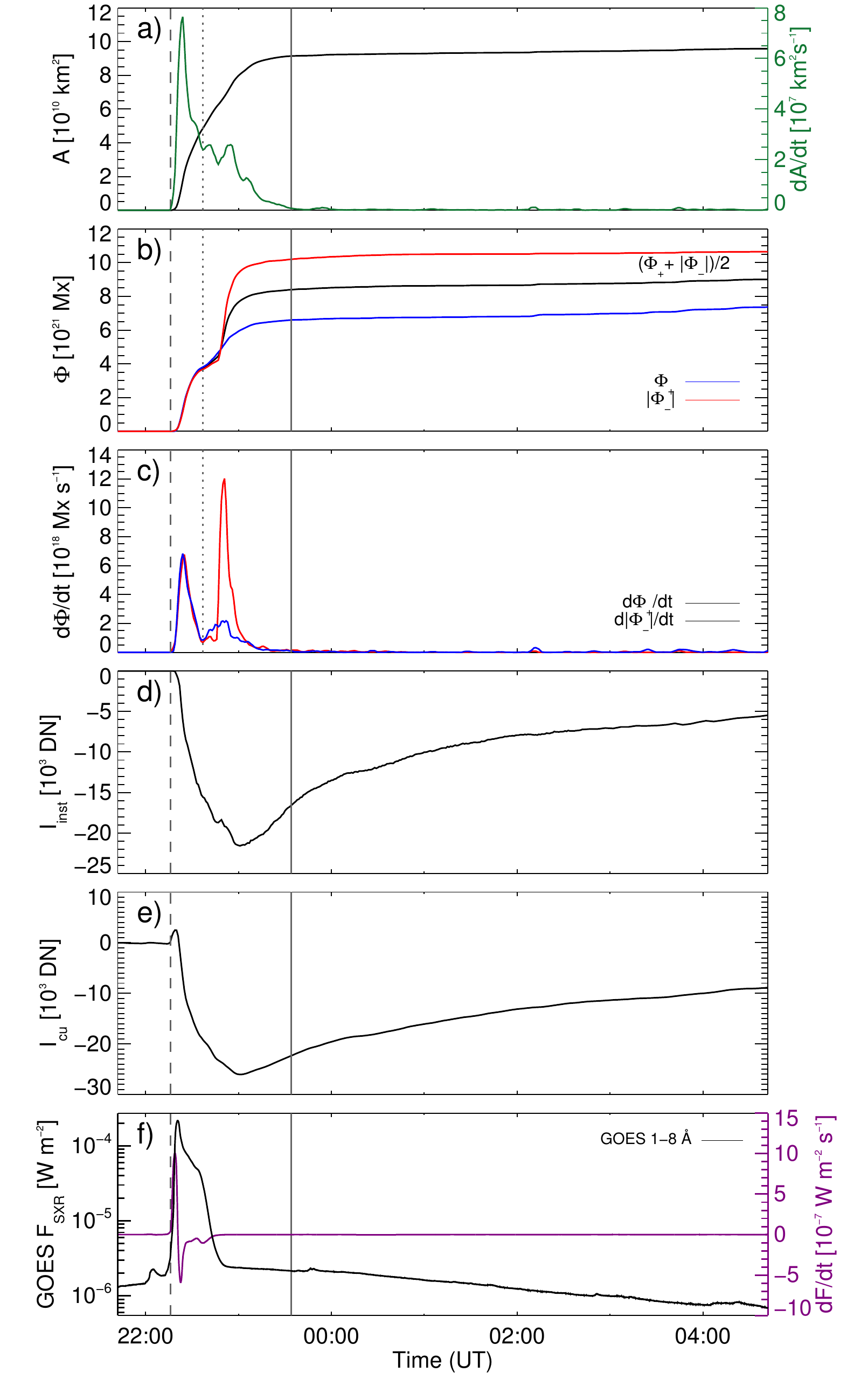}
\caption{Time evolution of coronal dimming parameters for September 6, 2011. From top to bottom we plot (a) the dimming area $A$ (black) and its time derivative, the area growth rate $\dot{A}$ (green), (b) the positive $\Phi_{+}$ (blue), negative $\Phi_{-}$ (red) and total unsigned magnetic flux $\Phi$ (black), (c) the corresponding magnetic flux rates $\dot{\Phi}_{+}$ (blue), $\dot{\Phi}_{-}$ (red), (d) the instantaneous total brightness $I_{inst}$, (e) the cumulative total brightness $I_{cu}$, and (f) the GOES 1.0--8.0 \AA~SXR flux (black) and its time derivative (purple). The vertical dashed line marks the start of the associated flare, the vertical dotted line outlines the end of the flux balance in the dimming, and the vertical solid line indicates the end of the impulsive phase of the dimming, respectively.}
\label{fig:profiles}
\end{figure}

The dimming region is expanding towards the North (cf.~Fig.~\ref{fig:detection} and \ref{fig:multi-wavelength}), following the direction of the associated EUV wave. \cite{Dissauer:2016} showed that this fast propagating structure coincides with the projected front of the CME. The true EUV wave has not yet decoupled from its driver at this time.
Our method of extracting the dimming region locates successfully the footpoints of the erupting flux rope, which are also identified in direct observations (cf.~Fig.~\ref{fig:core}).
Figure~\ref{fig:profiles} shows from top to bottom, the time evolution of selected dimming parameters derived from SDO/AIA \mbox{211 \AA}~together with the associated flare evolution. In panel (a) the dimming area $A$ (black line) and its time derivative, the area growth rate $\dot{A}$ (indicated in green) are plotted. The significant area growth starts almost co-temporal with the flare onset (cf.~panel f, vertical dashed line) and lasts for about two hours. We define this time range to be the \textit{impulsive phase} of the dimming. 
The area growth rate reaches its maximum value of \mbox{$7.7\times 10^{7}$~km$^{2}$~s$^{-1}$} only 8 minutes after the dimming started to form.
At the end of the impulsive phase (\mbox{$\sim$ 23:34~UT}, vertical solid line in Fig.~\ref{fig:profiles}), we find a total dimming area of about \mbox{$9.1\times 10^{10}$ km$^{2}$}.
The evolution of the magnetic fluxes over time is shown in panel (b). For the total unsigned magnetic flux $\Phi$ (indicated in black) we find \mbox{$8.4\times 10^{21}\pm 4.8\times 10^{20}$~Mx}, which is imbalanced and results from \mbox{$1.0\times 10^{22}\pm 5.0\times 10^{20}$~Mx} negative magnetic flux $\Phi_{-}$ (red line) and \mbox{$6.6\times 10^{21} \pm 4.8\times 10^{20}$~Mx} positive magnetic flux $\Phi_{+}$ (blue line), respectively. As can be seen from the time evolution of the corresponding magnetic flux change rates (see panel c), the same amount of positive and negative magnetic flux is involved in the early stage of the impulsive phase (until \mbox{$\sim$ 22:37~UT}). 
All dimming pixels that are identified during this flux balance period are shown in panel (b) of Fig.~\ref{fig:mask_timing}. Its comparison to the timing map of the overall dimming region (panel d) reveals that all pixels that are identified afterwards are located in regions around the negative polarity and an isolated region in the North-East. These regions mainly contribute to the secondary peak in the negative magnetic flux rate (Fig.~\ref{fig:profiles}, panel c, red curve) in the end of the impulsive phase.
At the end of the flux balance period (indicated as a vertical dotted line in Fig.~\ref{fig:profiles}) the total unsigned magnetic flux $\Phi$ amounts to \mbox{$3.75\times 10^{21}$~Mx} (\mbox{$\Phi_{+}=3.81\times 10^{21}$~Mx}, \mbox{$\Phi_{-}=3.68\times 10^{21}$~Mx}) and the dimming region covers an area of \mbox{$4.9\times 10^{10}$~km$^{2}$}. Furthermore, about 30\% of this balanced magnetic flux results from the localized core dimmings (cf.~Fig.~\ref{fig:core}, $\Phi=$\mbox{$1.1\times 10^{21}$}~Mx, $\Phi_{+}=$\mbox{$1.2\times 10^{21}$}~Mx, $\Phi_{-}=$\mbox{$9.9\times 10^{20}$}~Mx) that cover only $\sim$ 7\% of the balanced total dimming region.

The total brightness profiles $I_{inst}$ and $I_{cu}$ (cf.~Fig.~\ref{fig:profiles} d-e) show a similar behavior and reach their minimum at the same time. This indicates that the maximum dimming area is reached simultaneously with the total minimum intensity of the dimming pixels. This finding is supported by Fig.~\ref{fig:minimum_brightness_timing} where the majority of pixels within the cumulative dimming mask reach their minimum within the first two hours of the time series. 
\begin{figure}
\centering
\includegraphics[width=0.95\columnwidth]{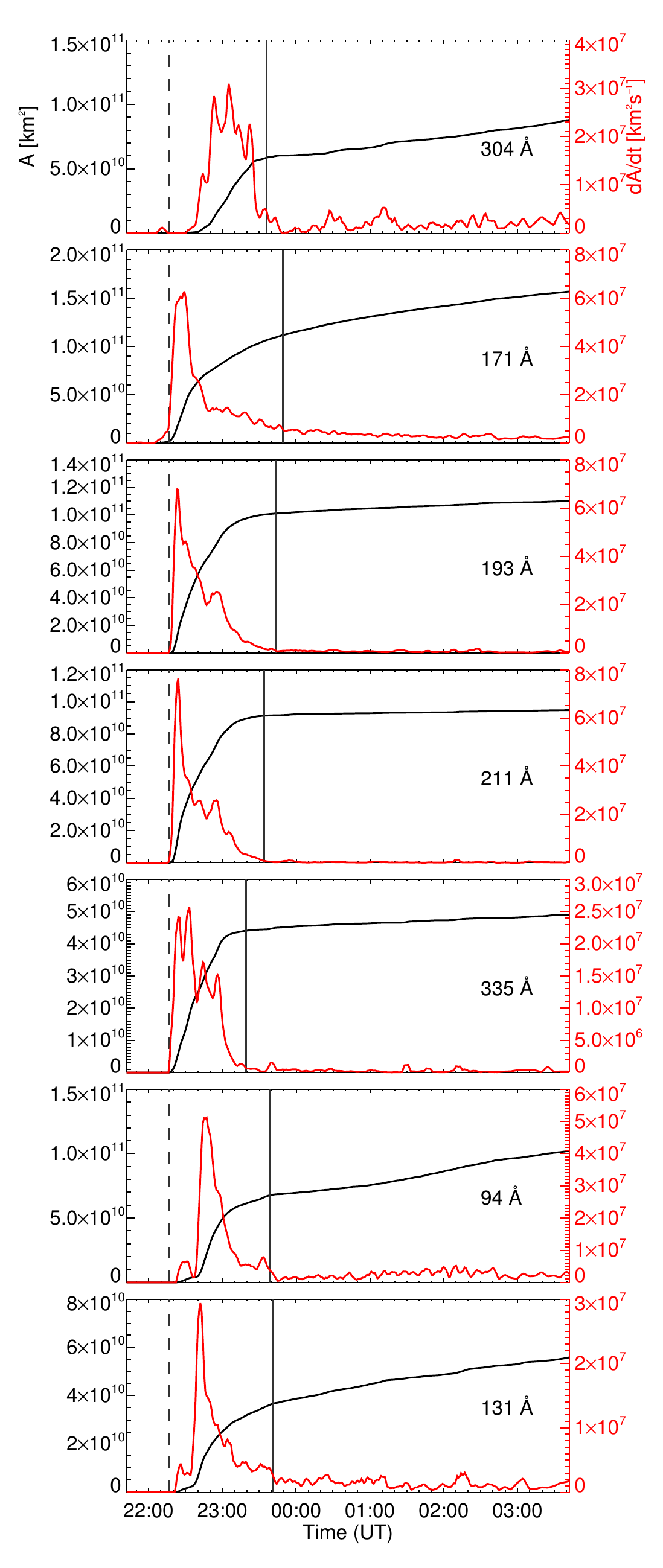}
\caption{Time evolution of the dimming area and the area growth rate (red line) for all SDO/AIA filters under study for September 6, 2011. The vertical dashed line indicates the start of the associated flare, while the vertical solid line marks the end of the impulsive phase of the dimming, respectively.}
\label{fig:multi_wave_area}
\end{figure}

Figure~\ref{fig:multi_wave_area} shows the time evolution of the dimming area and its growth rate for all seven SDO/AIA wavelengths under study. While the dimming is observed in all wavelengths, it is most pronounced in the 211, 193 and \mbox{171 \AA}~channels, indicating that mostly plasma at quiet coronal temperatures is evacuated. The values for the dimming area range from \mbox{$4.4\times10^{10} - 1.1\times10^{11}$~km$^{2}$} and for the area growth rate from \mbox{$2.6\times 10^{7} - 7.7\times 10^{7}$~km$^{2}$~s$^{-1}$} for the different channels. The impulsive phase of the dimming, i.e.~the time range during which the dimming area rapidly grows, starts co-temporal for 171, 193, 211 and \mbox{335 \AA}, which are all channels sensitive to quiet Sun coronal temperatures. For the 94, 131 and \mbox{304 \AA}~filters the area growth starts later (cf.~Fig.~\ref{fig:multi_wave_area}). 
94 and \mbox{131 \AA}~are sensitive to hot plasma of active regions, so in the beginning the flare dominates the emission and coronal dimming regions may not be identified yet. The \mbox{304 \AA}~filter represents plasma in the transition region and chromosphere. It is possible that plasma is evacuated in stages and that it simply takes longer to evacuate plasma from such low heights or that also signatures of the flare are present. 
\subsection{26 December 2011}
The dimming event occurred on December 26, 2011 in association with a partial halo CME observed by SOHO/LASCO with an average speed of about \mbox{v=740~km~s$^{-1}$}. The associated C5.7 flare started at \mbox{11:16~UT} with heliographic coordinates \mbox{N18\degree W02\degree}~in NOAA active region 11384. The GOES SXR emission reaches its peak around \mbox{11:50~UT}. A large-scale EUV wave is propagating to the North-East with a speed of around \mbox{v=680~km~s$^{-1}$} \citep{Nitta:2013}. The evolution of the dimming region related to the associated flare and CME has already been studied in detail in \cite{Cheng:2016} and \cite{Qiu:2017}. The unprojected maximal speed of the CME, as derived from STEREO-B, is about \mbox{$\sim 1100$~km~s$^{-1}$}.
\begin{figure*}
\centering
\includegraphics[width=0.9\textwidth]{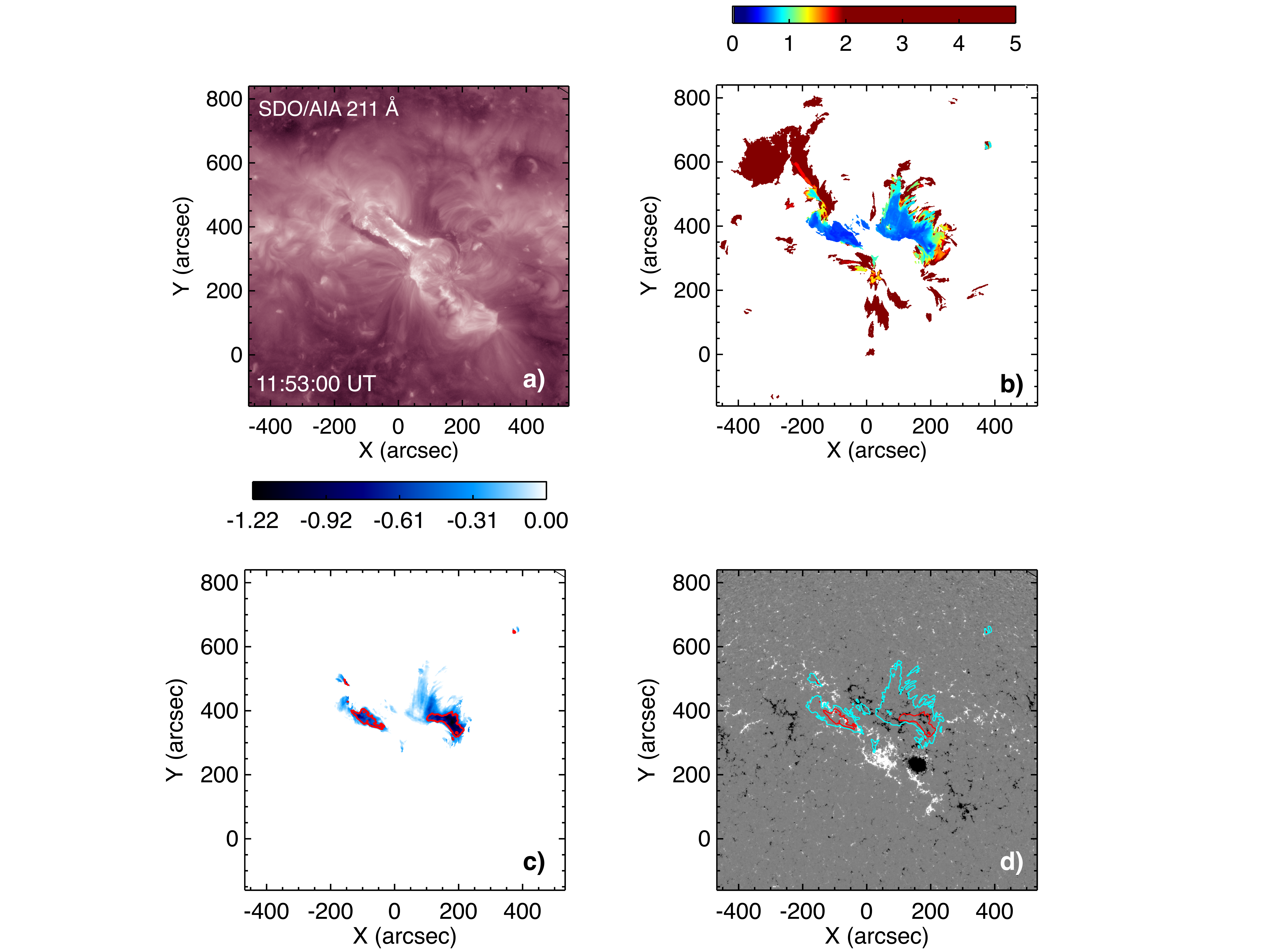}
\caption{SDO/AIA \mbox{211\AA}~direct image close to the time of maximum extent of the dimming (a), timing map indicating when each dimming pixel was detected for the first time (b), minimum intensity map of logarithmic base-ratio data (c) and SDO/HMI LOS magnetogram (d) illustrating the evolution of the coronal dimming that occurred on December 26, 2011. The identified potential core dimming regions are marked by the red contours. The cyan contours indicate the total dimming region identified during the impulsive phase.}
\label{fig:20111226_1046_results}
\end{figure*}

Figure \ref{fig:20111226_1046_results} gives an overview of the development of the dimming region.
Panel (a) shows a SDO/AIA \mbox{211 \AA}~direct image close to the maximum extent of the area. The dimming regions are compact and located next to the flare ribbons, in opposite polarity regions (cf.~panel d, cyan contours). This is supported by the temporal evolution of the dimming region (panel b), where blue and red correspond to the early and later stages of the dimming evolution and the colors in between resolve the main impulsive phase. Next to the main dimming region that is formed \mbox{30--60~minutes} after the flare start, another dimming region in the North-East is formed (indicated by the red pixels). 
The minimum intensity map plotted in panel (c) shows that the darkest pixels within the dimming region are located close to the eruption site (center of FOV of the image). The regions identified as core dimmings (red contours in panels c-d) coincide with the left and the right feet of the magnetic flux rope identified by \cite{Qiu:2017}.

\begin{figure}
\includegraphics[width=1.0\columnwidth]{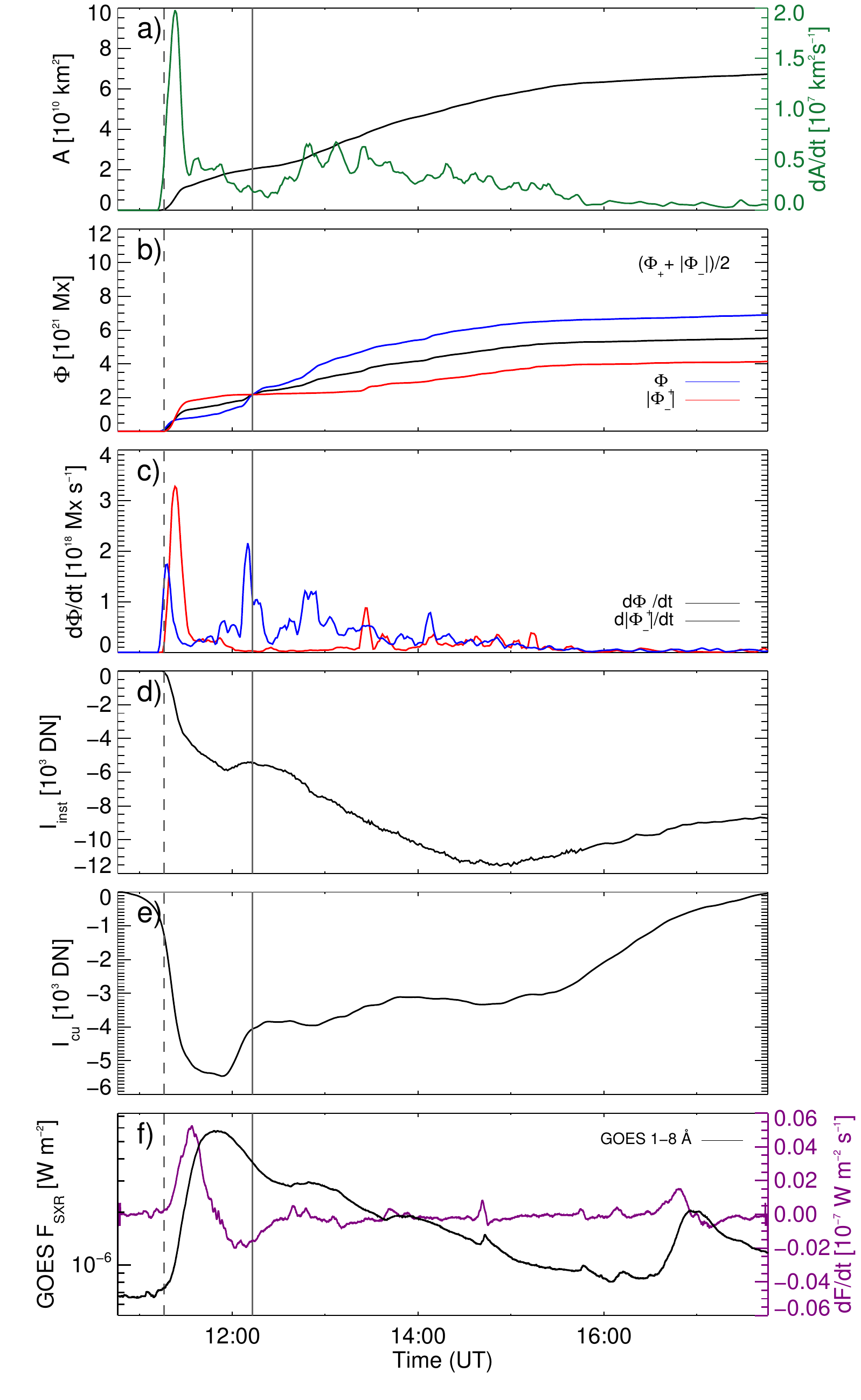}
\caption{Time evolution of coronal dimming parameters for December 26, 2011. From top to bottom we plot (a) the dimming area $A$ (black) and its time derivative, the area growth rate $\dot{A}$ (green), (b) the positive (blue), negative (red) and total unsigned magnetic flux (black), (c) the corresponding magnetic flux rates $\dot{\Phi}_{+}$ (blue), $\dot{\Phi}_{-}$ (red), (d) the instantaneous total brightness $I_{inst}$, (e) the cumulative total brightness $I_{cu}$ and (f) the GOES 1.0--8.0 \AA~SXR flux (black) and its time derivative (purple). The vertical dashed line indicates the start of the associated flare, while the vertical solid line marks the end of the impulsive phase of the dimming, respectively.}
\label{fig:2011226_1046_profiles}
\end{figure}

The time evolution of the extracted dimming parameters is plotted in Fig.~\ref{fig:2011226_1046_profiles}. Panel (a) shows the dimming area and its area growth rate. 
The impulsive phase of the dimming starts around \mbox{11:11~UT}, which is co-temporal with the start of the flare ribbon brightening and the fast rise of the CME \citep{Cheng:2016}. The maximum area growth rate of \mbox{$2.0 \times 10^{7}$~km$^{2}$~s$^{-1}$} is reached 12 minutes after the onset. The impulsive phase of the dimming stops after 60 minutes, around \mbox{12:13~UT}. The total dimming area amounts to \mbox{$2.0\times 10^{10}$~km$^{2}$}. Dimming pixels that are identified later, i.e.~that correspond to the second peak in the time evolution of the area growth rate, are not connected to the main dimming regions (cf.~Fig.~\ref{fig:20111226_1046_results} panel b). These pixels also lead to the magnetic flux imbalance that can be identified after the end of the impulsive phase (cf.~Fig.~\ref{fig:2011226_1046_profiles} panel c).
At the end of the impulsive phase, the total unsigned magnetic flux $\Phi$ of about \mbox{$2.3\times 10^{21}\pm 2.3\times 10^{20}$~Mx} is balanced ($\Phi_{+}=2.4\times 10^{21}\pm 3.3\times 10^{20}$~Mx, $\Phi_{-}=2.2\times 10^{21}\pm 1.5\times 10^{20}$~Mx; panels b-c). This value is higher compared to the magnetic flux obtained by \citep{Qiu:2017} (\mbox{$\sim 9.75\times 10^{19}$~Mx}) but reasonable since we estimate the flux of the total dimming region identified over time.
The total reconnection flux obtained from the time evolution of the flare ribbons amounts to a similar value of \mbox{$\sim 1.8\times 10^{21}$~Mx} \citep{Cheng:2016}. In a recent study by \cite{Temmer:2017} also similar values (up to a factor of 2) for the total reconnection flux and the total unsigned magnetic flux were found.

In addition, we find that around 22\% of the total unsigned magnetic flux results from the core dimming regions ($\Phi=$\mbox{$5.0\times 10^{20}$~Mx}, $\Phi_{+}=$\mbox{$3.8\times 10^{20}$~Mx}, $\Phi_{-}$=\mbox{$6.3\times 10^{20}$~Mx}) that cover 17\% of the total dimming region for this event.

The instantaneous total brightness $I_{inst}$ (panel d) reaches its minimum around \mbox{15:00~UT}, due to the presence of the dimming region that occurred later. The cumulative total brightness $I_{cu}$ (panel e) shows its minimum three hours earlier, taking into account only pixels of the primary dimming region.
\section{Discussion and Conclusion}\label{sec:discussion}
A new approach for the detection of coronal dimming regions in seven different EUV channels of SDO/AIA is presented and used for the analysis of two sample events. A thresholding algorithm is applied to logarithmic base-ratio images to detect dimming regions. 
This is beneficial due to the following reasons: 1) the intensity change in high-intensity and low-intensity regions is equally considered by using relative changes. We are therefore able to detect also coronal dimmings expanding towards quiet Sun region in the most objective way. 2) The logarithm allows us to equally quantify intensity changes over several orders of magnitude.

In contrast to other methods \citep[e.g.][]{Podladchikova:2005,Kraaikamp:2015} we do not restrict our dimming detection to the darkest intensity regions only. Coronal dimmings can be patchy in their appearance and not necessarily be represented by simply connected areas around the darkest pixel. We would miss parts of the dimming that is not connected to the darkest regions, especially for secondary dimmings. 
In addition to the total dimming region, our approach is also able to extract potential core dimming regions. They show the strongest decrease in intensity, lie close to the active region and in opposite magnetic polarity regions (see Fig.~\ref{fig:core}, \ref{fig:20111226_1046_results}).
Coronal dimming regions are in general complex, and different parts may grow and recover on different timescales. By introducing cumulative dimming pixel masks we combine all dimming pixels identified over a particular time range and are therefore able to capture the full extent of their area. This includes also regions initially obscured by the flare emission or post-flare loops.
Calculating characteristic dimming parameter, such as the dimming area and its magnetic flux from cumulative dimming pixels masks basically means that we extract time-integrated quantities. This is clearly in contrast to parameters extracted only at a certain time step \citep[e.g. at the maximum extent of the dimming region,][]{Reinard:2008}.
Furthermore, we are able to investigate the dynamical properties of coronal dimming regions, such as the area growth rate and the magnetic flux change rates.

We applied the new detection method to two events and characterized their dimming region by the time evolution of the newly defined dimming parameters. For both events we successfully identified core dimming regions, i.e. the footpoints of the erupting flux ropes. The overall dimming region expands around these footpoint regions further away from the eruption site (cf.~Fig.~\ref{fig:mask_timing} and Fig.~\ref{fig:20111226_1046_results}). For this early impulsive expansion phase, the magnetic flux involved in the dimming region is balanced and up to 30\% of it results from the localized core dimming regions.

Although the deprojected maximum speeds of the CMEs in the coronagraphic FOV are similar for both events (around \mbox{$\sim$ 1100--1200~km~s$^{-1}$}), the parameter values characterizing the dimming expansion are quite different. The dimming event observed on 6 September 2011 reaches a four times higher maximal area growth rate compared to the dimming that occurred on 26 December 2011. Also the duration from the dimming onset to its maximum area growth is shorter (\mbox{8~min} compared to \mbox{12~min}). 
CMEs reach on average their maximum speed at a distance of about \mbox{1.6~Rs}, while the maximum acceleration peak is found below \mbox{$< 0.5$~Rs} \citep{Temmer:2010,Bein:2011}. Furthermore, it was shown that the acceleration profiles of CMEs and the energy release of the associated flare evolve in a synchronized manner \citep{Temmer:2008,Temmer:2010,Berkebile-Stoiser:2012}.
It is therefore possible that dimming properties reflect more the initial driving and acceleration phase of the CME than the speed reached at the end of the impulsive acceleration phase \citep{Zhang:2001}.

The start of the impulsive area growth rate of the dimming is co-temporal to the flare onset for both events, and also to the rise time of the CME \citep[for 26 December 2011, see~][]{Cheng:2016}, further indicating a strong relationship of coronal dimmings with both, CMEs and flares. 
Differences in the size of the dimming area and therefore also in the magnitude of magnetic fluxes is furthermore more indicative for different CME masses. 
The detection method developed allows us to systematically analyze coronal dimmings, and to characterize their main physical properties and their evolution. In a future study, this method will be applied to a statistical set of coronal dimming events in optimized multi-point observations, where the dimming is observed against the solar disk by SDO/AIA and the associated CME evolution close to the limb by STEREO/EUVI, COR1 and COR2.
\section{Acknowledgements} 
We gratefully acknowledge the support by the Austrian Space Applications Programme of the Austrian Research Promotion Agency FFG (ASAP-11 4900217) and the Austrian Science Fund FWF (P24092-N16). We thank the anonymous referee for constructive comments which helped to improve the manuscript. {\it SDO} data are courtesy of the NASA/{\it SDO} AIA and HMI science teams.
\bibliographystyle{aasjournal}
\bibliography{biblio} 
\end{document}